# Strain controlled superconductivity in few-layer NbSe$_2$

Cliff Chen[1], Protik Das[2], Ece Aytan[2], Weimin Zhou[1], Justin Horowitz[1], Biswarup Satpati[3], Alexander A. Balandin[2], Roger K. Lake[2] and Peng Wei[1]*

1. Department of Physics and Astronomy, University of California, Riverside, CA 92521, United States

2. Department of Electrical and Computer Engineering, University of California, Riverside, CA 92521, United States

3. Surface Physics & Material Science Division, Saha Institute of Nuclear Physics, 1/AF Bidhannagar, Kolkata 700 064, India

*Correspondence to: peng.wei@ucr.edu

**Abstract**: The controlled tunability of superconductivity in low-dimensional materials may enable new quantum devices. Particularly in triplet or topological superconductors, tunneling devices such as Josephson junctions etc. can demonstrate exotic functionalities.[1] The tunnel barrier, an insulating or normal material layer separating two superconductors, is a key component for the junctions. Thin layers of NbSe$_2$ have been shown as a superconductor with strong spin orbit coupling,[2-4] which can give rise to topological superconductivity if driven by a large magnetic exchange field.[5-9] Here we demonstrate the superconductor-insulator transitions in epitaxially grown few-layer NbSe$_2$ with wafer-scale uniformity on insulating substrates. We provide the electrical transport, Raman spectroscopy, cross-sectional transmission electron microscopy, and X-ray diffraction characterizations of the insulating phase. We show that the superconductor-insulator transition is driven by strain, which also causes characteristic energy shifts of the Raman modes. Our observation paves the way for high quality heterojunction tunnel barriers to be seamlessly built into epitaxial NbSe$_2$ itself, thereby enabling highly scalable tunneling devices for superconductor-based quantum electronics.

**I. Introduction**:

Imposing structural perturbations to thin layers of NbSe$_2$ can be used as a potential knob to control superconductivity. NbSe$_2$ is often found in its prismatic 2H phase, i.e. the superconducting phase, in devices fabricated from exfoliated thin crystals.[2,4,10] Insulating NbSe$_2$ is rare,[11] although it has been recently demonstrated in the 1T phase of NbSe$_2$.[12] However, a controlled tuning in between the superconducting and insulating states of NbSe$_2$ has not yet been achieved. NbSe$_2$ thin layers are often either exfoliated or grown on another two-dimensional (2D) material such as graphene.[3,13] Due to Van der Waals interaction, the coupling between the layer and the substrate is week, and the substrate often does not play a major role in determining the electrical properties of NbSe$_2$.

**II. Experimental Methods and Results**:

We demonstrate the electrical properties of few-layer NbSe$_2$ grown using MBE on a non 2D material substrate, i.e. on sapphire. Because the growth is epitaxial, the sapphire substrate couples to the NbSe$_2$ layer. One evidence of such coupling is seen from the locked [210] crystalline directions between the substrate and the NbSe$_2$ layer as observed in situ by reflection high energy electron diffraction (RHEED) during the growth (Fig 1c). Ideally, NbSe$_2$ has a much smaller lattice constant ($a \sim 3.44$ Å) [14] compared to that of the *c*-cut sapphire ($a \sim 4.76$ Å), and one would not expect the growth to be epitaxial. However, by considering the 3×3 unit cells of NbSe$_2$ and the 2×2 unit cells of the *c*-cut sapphires surface, the two crystals do match with each other and the mismatch is only ~ 7.8%. Such superlattice cell matching has been previously reported in the epitaxial growth of high quality transition-metal dichalcogenides (TMD) thin layers, for example WSe$_2$, on sapphire substrate.[15] Notably, when considering the superlattice matching, the *c*-cut sapphire substrate imposes a planar compressive strain to NbSe$_2$. Similar heterogeneous growth, i.e. growing a material epitaxially on top of another one with drastically different properties, has demonstrated many emerging physics phenomena. One example is the epitaxial growth of Fe on GaAs(001) giving rise to the Fe/GaAs(001) heterostructures, in which tunneling anisotropic magnetoresistance and spin-orbit proximity effects have been shown.[16,17]

By controlling the growth temperatures, we obtain the $NbSe_2$ layers with two distinctive phases: one being a superconductor and the other one being an insulator (or semiconductor). Because our substrate is sapphire, we are able to carry out the electrical transport characterizations of both phases. We confirm that the insulating (or semiconducting) $NbSe_2$ has a gapped Fermi surface and also demonstrates a shift of the Raman peaks with respect to the standard 2H-$NbSe_2$ Raman modes. It is known that the Raman peak shift often indicates a hardening or softening of the lattice vibration modes, which may result from lattice strain effect. Using selective area diffraction (SAD) based on cross-sectional transmission electron microscopy (TEM) and X-ray diffraction (XRD), we show a lateral compression and a vertical expansion of the $NbSe_2$ lattice - agreeing well with the Raman results. Our findings provide insight to the transition between the insulating and superconducting electronic properties of the two phases.

The quality of our epitaxially grown $NbSe_2$, including crystal symmetry and epitaxial orientation, is confirmed by in situ RHEED experiment (Fig. 1a-1c). A 0.5 monolayer (ML) $NbSe_2$ seed, which does not completely cover the sapphire substrate, is used to facilitate the growth (see SI). The six-fold rotational symmetry of the grown layer is confirmed by a planar rotation of the RHEED electron beam from [210] to [110] directions (Fig. 1c). It is also clear that the crystalline orientations of $NbSe_2$ and sapphire are locked as $[210]_{NbSe_2}$ // $[210]_{sapphire}$ (Fig. 1a-1c), which infers that there is an interaction between $NbSe_2$ and the substrate, and thus would lead to strain effect.

The growth is proved to be layer-by-layer. Each monolayer of NbSe$_2$ is clearly defined as shown in the TEM image (Fig 1d, 5ML NbSe$_2$). According to the scale bar of the TEM image, Fig 1d also shows that each NbSe$_2$ monolayer extends continuously over a large distance. It thus suggests that atomically thin NbSe$_2$ on sapphire is achievable using MBE. The wafer-scale coverage of NbSe$_2$ is further confirmed by both XRD and atomic force microscopy (AFM) (Fig. 2a and 2b). The quality of NbSe$_2$ is demonstrated by a set of characteristic XRD diffraction peaks (Fig. 2a). AFM shows that the grown NbSe$_2$ layer has a wafer-scale coverage over the substrate (Fig 2b). The terrace-like feature in Fig 2b indicates that the NbSe$_2$ layer follows the atomic terraces of sapphire, which agrees with the uniform island-free growth of NbSe$_2$. The crystalline quality of the NbSe$_2$ samples, down to a few MLs, is also confirmed by electron diffractions carried out locally in a cross-sectional region of NbSe$_2$ using SAD (Fig 2c). The NbSe$_2$ diffraction pattern is distinct from the substrate pattern with the (002) series diffraction spots aligned with the (006) series of the sapphire substrate. The spatially resolved energy dispersive X-ray spectroscopy (EDX) line scan shows no element segregations in the layer and a clear interface with the substrate. The EDX also shows no contamination in the NbSe$_2$ layer, and an atomic ratio ~ 1:2 for Nb:Se (Fig S2b in SI). Further, using atomically resolved scanning transmission electron microscopy (STEM), the relative arrangement of Nb and Se atoms is clearly observed (Fig S1c in SI). The above studies confirm that stochiometric NbSe$_2$ layers have been grown.

A systematic tuning of the superconductivity is achieved by controlling the growth condition of in NbSe$_2$ (Fig. 3a). Both increasing the growth temperature and decreasing the layer thickness cause a decrease of $T_C$ (Fig. 3b). However, as the growth temperature increases, even in thick samples, the sample resistance starts to develop a sharp upturn at low temperature (Fig. 3b inset). To clarify the origin of the resistance upturn, we increase the growth temperature to 600 °C, which gives rise to an insulating (or semiconducting) NbSe$_2$ sample in the full temperature range (Fig. 3a blue). A plot of $\ln(R)$ vs. $T^{-1}$ is used to validate the origin of the insulating behavior. We find that a linear fit well describes the $\ln(R)$ vs. $T^{-1}$ plot (Fig. 3a inset) in a wide temperature range (13 K and above), which is not the case if a variable range hopping model is adopted (Fig S4 in SI). Thus, the insulating behavior is not due to localization effects caused by defects. It confirms a gapped Fermi surface with the conductivity given by $\sigma \sim e^{-\frac{E_0}{k_B T}}$, where the activation energy determined to be $\frac{E_0}{k_B} \sim 6.7$ K. Interestingly, we see little changes on crystal structure between the superconducting and insulating samples (Fig S2a). Both have similar lattice constants and the same hexagonal rotation symmetry (Fig S2a). Further, the insulating sample (grown at $T_g$ ~ 600 °C) has even better sample quality (Fig S2a and Fig S3). Therefore, the insulating behavior is not due to the degradation of the NbSe$_2$ sample when growing at 600 °C.

Although RHEED does not show visible changes, the difference between the two types of NbSe$_2$ samples is visible in Raman spectroscopy, which is known to be sensitive to various structural transitions in a wide range of materials.[18,19] First, we compare the Raman spectrum of a 5 ML superconducting sample to that of a bulk NbSe$_2$ flake in Fig 3c. The two spectra resemble each other by both showing the A$_{1g}$, E$^1_{2g}$ and the soft mode peaks, which confirms the quality of the MBE grown layer. Compared to bulk NbSe$_2$, it is clear that the E$^1_{2g}$ peak of the MBE grown NbSe$_2$ blue shifts to a higher wave number, while the A$_{1g}$ peak red shifts to a lower wave number (Fig 3c). The nature of the E$^1_{2g}$ and A$_{1g}$ vibration modes is further confirmed by polarized Raman spectroscopy (Fig 3d), in which the *p*-polarization turns off the A$_{1g}$ peak while maintaining the E$^1_{2g}$ peak, consistent with prior reports in 2H-NbSe$_2$.[10] The blue shift of E$^1_{2g}$ or the red shift of A$_{1g}$ is related to the hardening or softening of phonons, which could be a result of the change of lattice parameters due to strain.[20] It is also known that the hardening of a phonon mode is often unfavorable to superconductivity.

To further understand the change of the Raman modes, we carried out a series of Raman spectroscopy experiments with a focus on the shift of the $E^1_{2g}$ and $A_{1g}$ peaks (Fig 4a). The measurement is done in a set of NbSe$_2$ samples (ranging from superconducting to insulating) that are grown at difference temperatures. The corresponding resistance as a function of temperature is plotted in Fig 3a. A two-peak Gaussian fit is adopted to identify the $E^1_{2g}$ and $A_{1g}$ peak locations (Fig 4a). A clear increase of the $E^1_{2g}$ peak, from 245 cm$^{-1}$ to 260 cm$^{-1}$, is observed, which corresponds to the transition from superconducting to insulating in NbSe$_2$. At the same time, the $A_{1g}$ peak decreases. The validity of the comparison is guaranteed by the aligned sapphire substrate Raman peak at 418 cm$^{-1}$. It has been shown that the $E^1_{2g}$ mode represents the vibration of atoms within a monolayer layer of NbSe$_2$, and the $A_{1g}$ mode represents the vibration perpendicular to the monolayer.[21] The opposite shift of the $E^1_{2g}$ and $A_{1g}$ modes could reflect different strain effects parallel and perpendicular to the monolayer. Because the (001) plane of NbSe$_2$ is in parallel to the (001) plane of sapphire, the shift of $E^1_{2g}$ could be due to the strain effect within the NbSe$_2$ monolayer. To confirm it, we carried our SAD in both the superconducting sample (growth temperature $T_g \sim 400$ °C) and the insulating sample (growth temperature $T_g \sim 600$ °C). Fig 4b shows a result of the comparison of SAD. There is a clear shift of the NbSe$_2$ ($\underline{1}$22) SAD diffraction spot. The validity of the comparison is based on the well aligned NbSe$_2$ (002) and Al$_2$O$_3$ ($\underline{1}$02) SAD diffraction spots by knowing that the NbSe$_2$ (002) spot is along the same direction as the Al$_2$O$_3$ (006) spot (Fig 2c). Comparing the insulating sample ($T_g \sim 600$ °C) with the superconducting sample ($T_g \sim 400$ °C), one can see that the NbSe$_2$ ($\underline{1}$22) SAD spot moves away in the *k*-space, which indicates a planar compression of the crystal lattice for the insulating sample.

Furthermore, to accurately determine the change of lattice constants along the *c*-axis direction, we carried out XRD measurements on both types of samples (Fig 4c). After aligning the Al$_2$O$_3$ (006) peaks, one can see that the insulating sample (T$_g$ ~ 600 °C) has a lower NbSe$_2$ (002) peak, which proves the increased (or expanded) *c*-axis lattice constant. The planar compression and out-of-plane expansion of the crystal lattice agrees with the Raman results, in which the E$^1_{2g}$ has a blue shift and A$_{1g}$ has a red shift (Fig 4a). The planar lattice compression agrees with the aforementioned lattice mismatch between NbSe$_2$ and the sapphire substrate when considering the 7.8% smaller sapphire supercell compared to that of NbSe$_2$, which has also been shown in the MBE growth of other TMD materials on sapphire.[15] On the other hand, the compression of the planar crystal lattice could also result in an increase of the *c*-axis lattice constant causing a red shift to the A$_{1g}$ mode that describes the out-of-plane lattice vibrations.

### III. Discussion:

Although we only measured the phonon modes at large wave numbers, such as the E$^1_{2g}$ and A$_{1g}$, when the lattice is under strain, the phonon modes at low wave numbers will also be modified. According to BCS theory,[22] the electron-phonon coupling strength is weakened when the stiffness of the lattice vibrations increases, which can be caused by a compressive strain. Similar effect has been previously reported in other 2D materials such as MoS$_2$.[23] Our observations are in agreement with them, and suggest that strain could cause the reduced T$_c$ and the switching-off of superconductivity in our metallic NbSe$_2$ samples (Fig 3b inset). Interestingly, we only see lattice compression within the monolayer plane, whereas the *c*-axis is expanded (Fig 4c). Therefore, our results point out that the planar vibration modes play a more important role on superconductivity in NbSe$_2$.

According to SAD (Fig 4b) and RHEED (Fig 4b and Fig S2c in SI) results, no structural phase differences are observed in between the superconducting and insulating samples. Previously, it has been reported that 1T-NbSe$_2$ is a Mott insulator, which is obtained by growing NbSe$_2$ at a higher substrate temperature.[12] Our density functional theory (DFT) calculations show that the Raman activated modes are very different between 2H-NbSe$_2$ and 1T-NbSe$_2$ (Fig S4 in SI). Thus, our Raman results in Fig 4a do not suggest that our insulating sample is 1T-NbSe$_2$. We infer that the insulating behavior could be due to the modified electron correlations such as the charge density wave (CDW) or the change of band structure in NbSe$_2$ as a result of strain, which is often seen in TMD materials.[24]

Our results show that strain can serve as an effective knob in tuning the superconducting properties of NbSe$_2$, which has been shown hard to be achieved by other means such as carrier density tuning.[25-27] The semiconducting (or insulating) NbSe$_2$ could serve as a high-quality tunnel barrier material with a good lattice match with the superconducting 2H-NbSe$_2$, thereby would enable a range of superconducting tunneling devices to be seamlessly built into NbSe$_2$. Further, the semiconducting (or insulating) NbSe$_2$ may allow gate-controlled insulator to superconductor transitions. TMD semiconductors in their 2H phase, for example 2H-MoS$_2$, can become superconducting once significant amount of charge carriers are injected into the material by gating.[28] Because our observed semiconducting (or insulating) NbSe$_2$ is in its 2H phase, superconductivity can potentially be turned on by a dielectric gate. It would thus enable switchable superconducting devices. These may also suggest the potential realization of 2D planar tunneling devices such as 2D Josephson junctions in NbSe$_2$, which has been predicted as a scalable platform for topological superconductivity and the manipulation of Majorana zero modes for fault-tolerant topological quantum computing.[29]

**IV. Conclusion**:

The growth of few-layer NbSe$_2$ using MBE has the advantage of producing heterostructures with ultra-clean interfaces, which are crucial for delivering high quality proximitized materials.[17] Because the grown NbSe$_2$ layers are very thin, interface effect dominates the electrical properties of the heterostructure. New phenomena due to various proximity effects, such as magnetic proximity effects, superconducting proximity effects and spin-orbit proximity effects, can emerge. For magnetic proximity effect, heterostructures between NbSe2 and other materials with large magnetic exchange fields[30,31] can be grown in-situ by MBE. The in-situ growth enables ultra-clean interface that is crucial for the induced exchange field, and such NbSe$_2$ heterostrucures are promising candidates for topological superconductors. For spin-orbit proximity effect, heterostructures can be grown by combining NbSe$_2$ with other materials having strong spin-orbit coupling (SOC), for example TMD materials WTe$_2$, WSe$_2$ etc. NbSe$_2$ can acquire stronger SOC or an additional interface SOC, which can be further tuned by a dielectric gate thereby enabling new spintronic devices.

**V. Methods:**

Growth of the NbSe$_2$ thin films was carried out in an ultrahigh vacuum chamber with a base pressure of ~ 5×10$^{-7}$ torr. The films were monitored in-situ throughout the growth by reflection high energy electron diffraction (RHEED) with a 7.5mW beam energy. Ex-situ Raman spectroscopy was carried out on uncapped films using a Horiba LabRam system. A 6mW, 532nm unpolarized excitation laser with a 100μm spot size was used to scan the films. Ex-situ x-ray diffraction was done using a PANalytical Empyrean Series 2 diffractometer with Cu K-Alpha1 line. Transport measurements were carried out in a homemade liquid helium probe. The sample was mounted inside the probe using pressed indium contacts, purged with He gas, and then cooled in a liquid helium dewar. Temperature was controlled by adjusting the insertion height of the probe. The temperatures below 4.2K were achieved by letting additional He gas condense inside the probe followed by pumping to reduce the vapor pressure. A Stanford Research Systems lock-in amplifier was used to measure the resistance of the films in a four-terminal configuration. The DFT calculations were carried out within the Perdew-Burke-Ernzerhof (PBE) generalized gradient approximation and the projected augmented wave (PAW) method as implemented in software package VASP.[32,33] The van der Waals (vdW) interactions were accounted for using Grimme's DFT-D2 semi-empirical correction to the Kohn Sham energies.[34]

**Acknowledgments:**

We acknowledge Krassimir N. Bozhilov and the Central Facility for Advanced Microscopy and Microanalysis (CFAMM) at UCR for technical supports. The research is supported as part of the Spins and Heat in Nanoscale Electronic Systems (SHINES), an Energy Frontier Research Center funded by the U.S. Department of Energy (DOE), Office of Science, Basic Energy Sciences (BES), under Award # SC0012670 (A.A.B., R.K.L. and P.W.). C.C., W.Z., J.H. and P.W. acknowledge the support of the startup fund from UC Riverside, and partly by the National Science Foundation (NSF) QLCI-CG under Award #1937155. The DFT calculations used the Extreme Science and Engineering Discovery Environment (XSEDE),[35] which is supported by the National Science Foundation Grant No. ACI-1548562 and allocation ID TG-DMR130081.

**Figures and captions:**

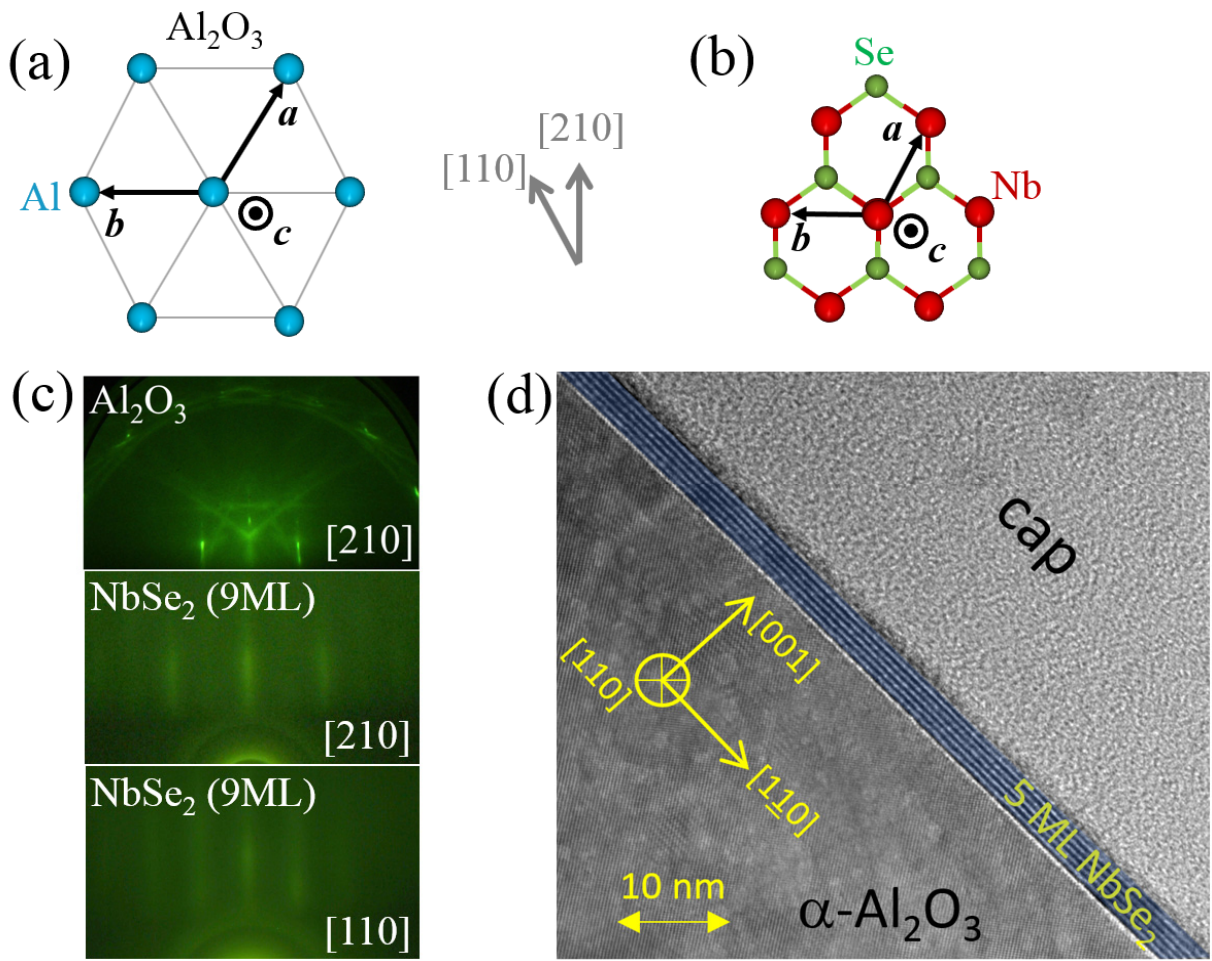

**Fig. 1.** (a) The schematic showing the Al atoms on sapphire (0001) surface. (b) The schematic of NbSe$_2$ surface and its epitaxial orientation with respect to the substrate in (a). (c) The RHEED patterns: sapphire substrate along [210] direction (top), 9 ML NbSe$_2$ along [210] direction (middle) and along [110] direction (bottom). (d) TEM image of a 5 ML NbSe$_2$ sample. It can be clearly seen that each NbSe$_2$ monolayer extends continuously over a large scale.

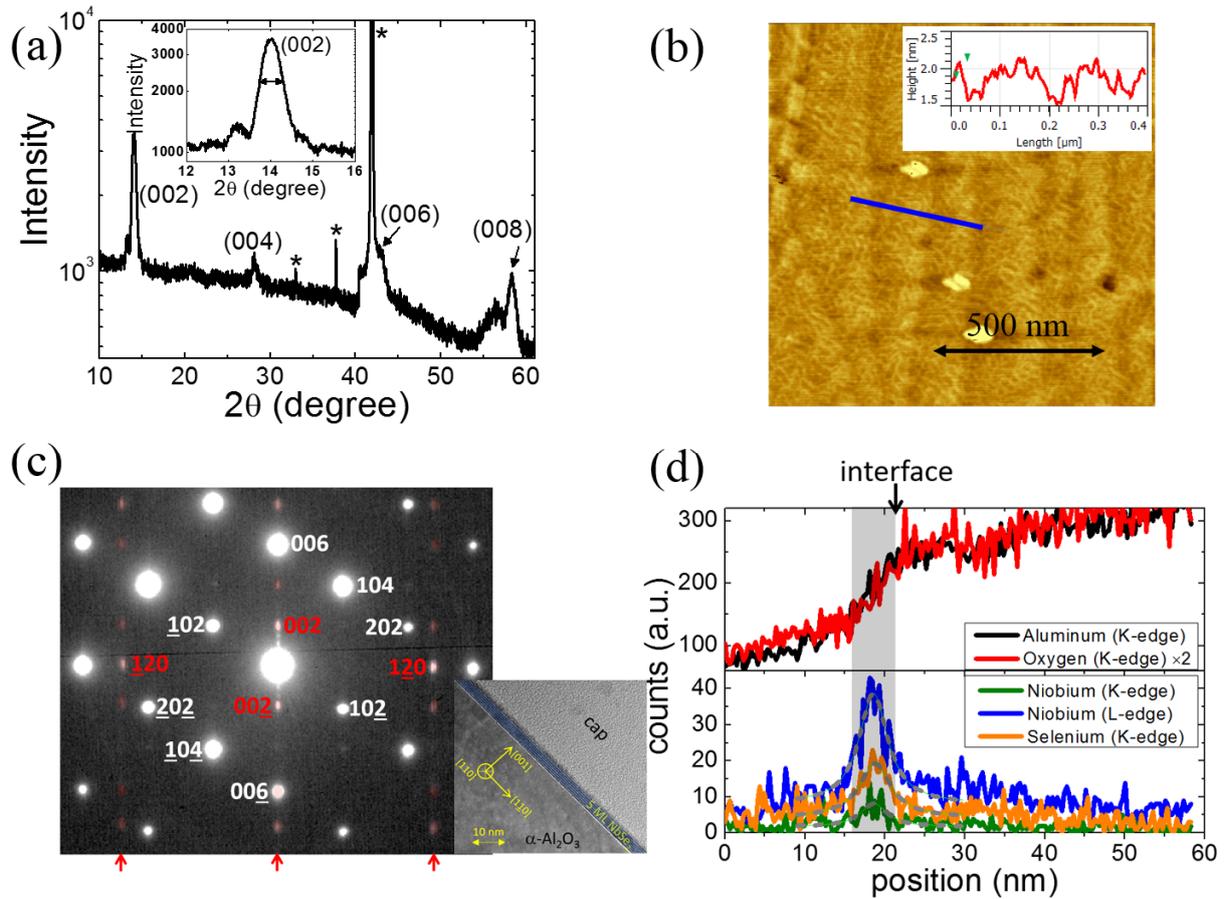

**Fig. 2.** (a) The X-ray diffraction of a NbSe$_2$ sample. The well oriented *c*-axis diffraction peaks are seen. The starred peaks correspond to the substrate and the background from the instrument. The inset provides a magnified image of the (002) peak, which also shows signatures of thickness fringes. (b) The AFM image of a continuous (islands-free) 3 ML NbSe$_2$ sample following the growth procedure shown in SI. The substrate terraces are visible indicating good surface coverage. (c) The TEM based selective area diffraction (SAD) data of the 5 ML NbSe$_2$ sample (inset). The substrate diffraction peaks are labeled in white and the NbSe$_2$ diffraction peaks are in red. The *c*-axis (002) diffraction of NbSe$_2$ is aligned with the *c*-axis (006) diffraction of sapphire. (d) The element selective spatially resolved energy dispersive X-ray spectroscopy (EDX) data of a line across the interface of the layer. The data shows no element segregations in the layer and a clear

interface with the substrate. The stoichiometric ratio between Nb and Se is determined in Fig S2b in the SI.

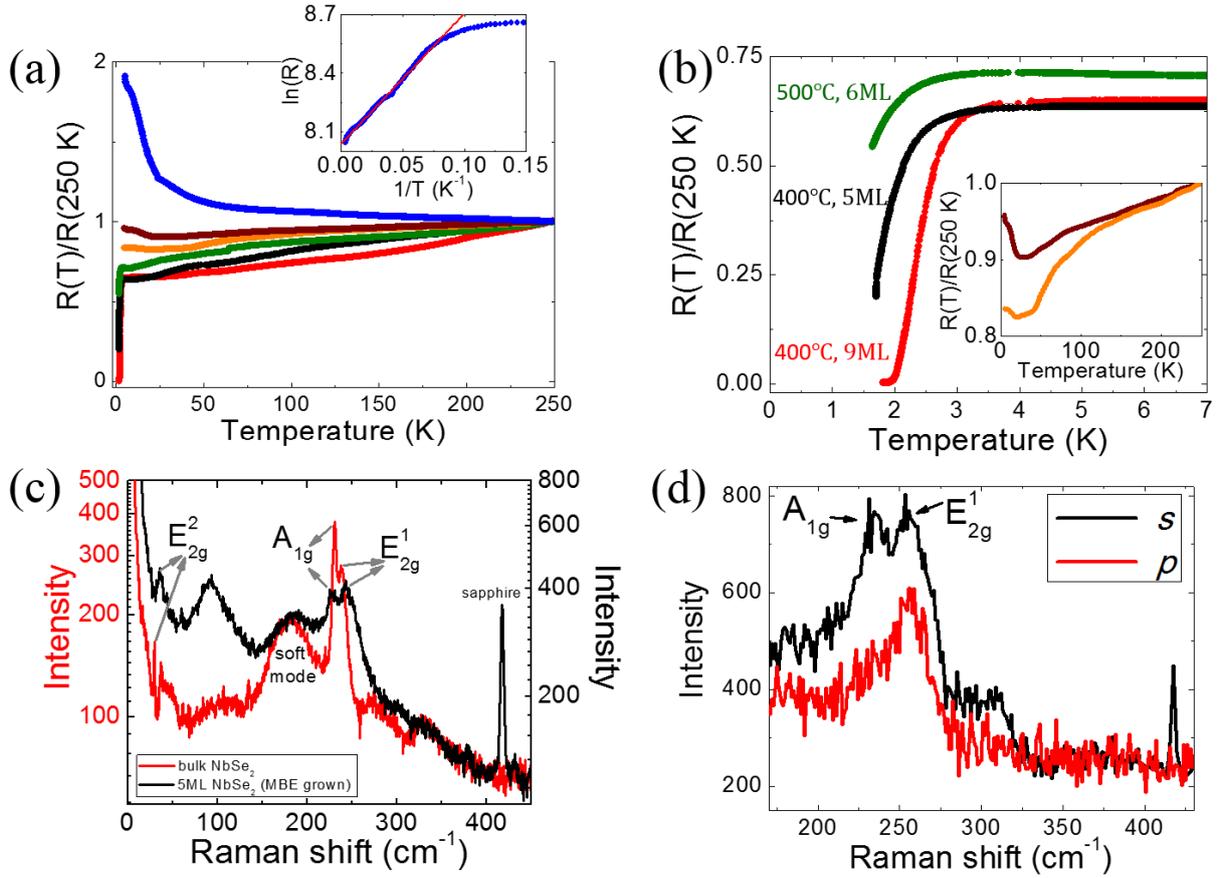

**Fig. 3.** (a) Resistance measurements of the superconductor to insulator transition in NbSe$_2$ samples systematically tuned by varying the growth temperature (from 400 °C to 600 °C). Inset: the ln(R) vs. 1/T plot of the insulating sample. The linear fit works in the full temperature range from 13 K and above, which gives an excitation gap $\frac{E_0}{k_B} \sim 6.7$ K. (b) The low temperature zoomed-in plot of the superconducting samples. Inset: the metallic samples demonstrating low temperature insulating behavior. The color of each curve matches that of (a). (c) Raman spectra comparison between MBE grown superconducting NbSe$_2$ (black) and a bulk sample (red). The MBE grown sample has a blue shift for E$^1_{2g}$ and a red shift for A$_{1g}$. (d) Polarized Raman spectroscopy confirming both the A$_{1g}$ and E$^1_{2g}$ Raman modes of the sample shown in the middle panel of Fig 4a. The A$_{1g}$ is turned off when switching from *s*- to *p*-polarization, whereas the E$^1_{2g}$ peak survives.

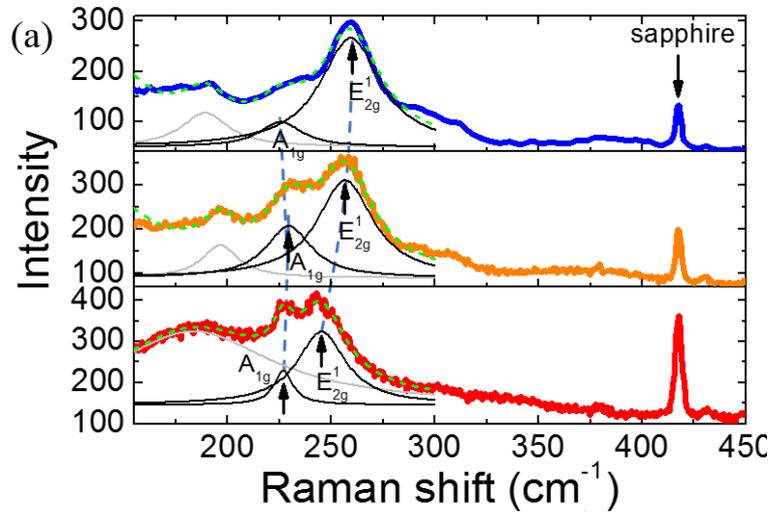

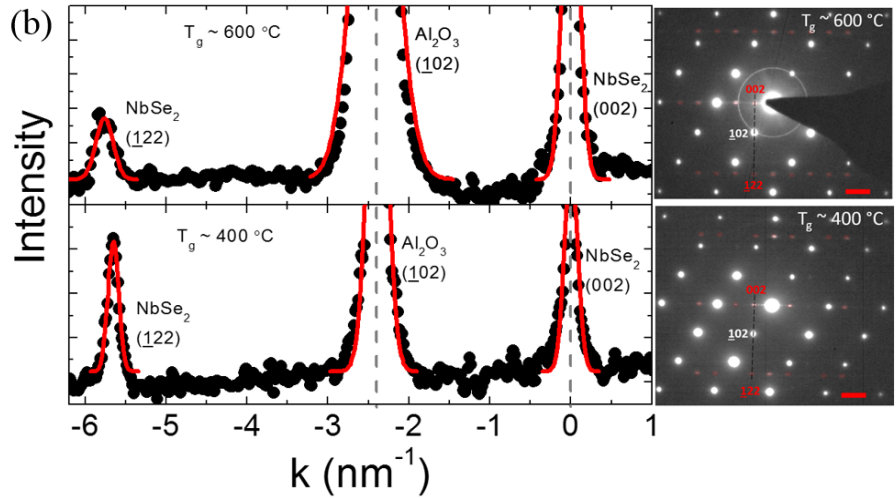

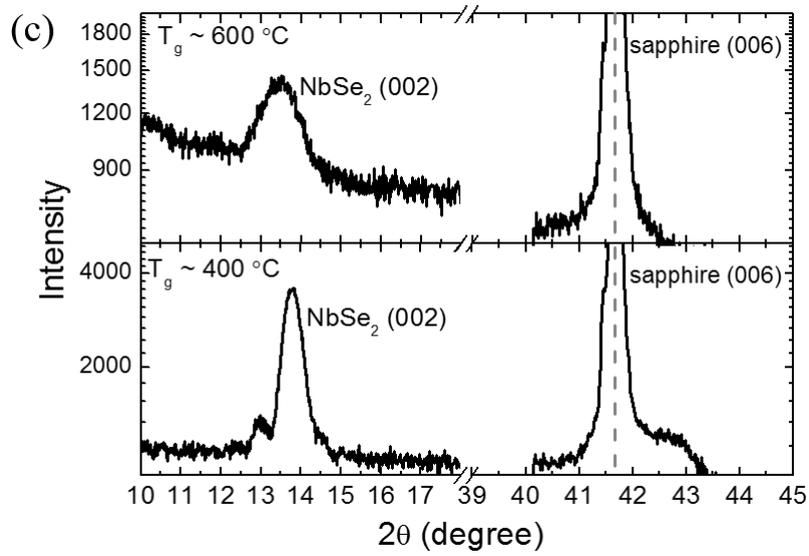

**Fig. 4.** (a) A comparison of the Raman spectra among three samples having different electrical properties (the superconducting sample, the insulating sample and a sample in between). The color of the data matches that in Fig. 3a. Two-peak Gaussian fit is used to identify the locations of the $E^1_{2g}$ and $A_{1g}$ peaks. A red shift of the $A_{1g}$ peak and a blue shift of the $E^1_{2g}$ peak are observed, which accompanies the loss of superconductivity. The comparison is guaranteed by the aligned sapphire peak. (b) The comparison of the SAD results of two samples grown at $T_g \sim 600\ °C$ (top, insulating) and at $T_g \sim 400\ °C$ (bottom, superconducting) respectively. A line cut is made from the NbSe$_2$ SAD (002) peak to the NbSe2 SAD ($\underline{1}22$) peak crossing the sapphire SAD ($\underline{1}02$) peak. The line cut is indicated by the dashed line shown in the TEM images (right). Gaussian fit is used to identify the locations of the diffraction peak. By aligning the NbSe$_2$ (002) and sapphire ($\underline{1}02$) peaks (left), one can see a clear shift of NbSe$_2$ ($\underline{1}22$) peak indicating that the sample grown at $T_g \sim 600\ °C$ has a slightly smaller lattice constant. (c) The XRD results of two samples grown at $T_g \sim 600\ °C$ (top) and at $T_g \sim 400\ °C$ (bottom). The NbSe$_2$ (002) XRD peak shifts to lower angel for the sample grown at $T_g \sim 600\ °C$ suggesting an expanded *c*-axis lattice constant. The comparison is based on the aligned sapphire (006) XRD peak.

**Table of Contents (TOC) graphic**

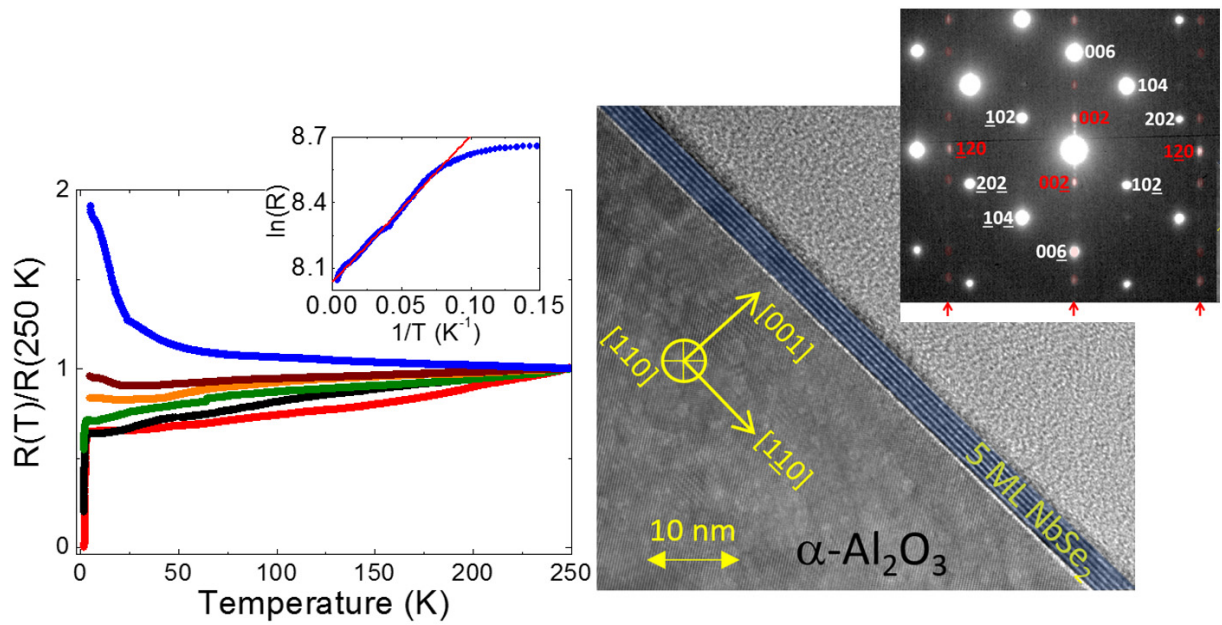

**Supporting Information**: including Sample growth (S1), Sample characterization (S2), Variable range hopping fitting to the insulating sample (S3), DFT calculations of the phonon spectrum of 1T-NbSe$_2$ (S4)

**References:**

1. Kitaev, A. Y. Unpaired Majorana fermions in quantum wires. *Phys-Usp+* **44**, 131, (2001).
2. Xi, X., Wang, Z., Zhao, W., Park, J.-H., Law, K. T., Berger, H., Forro, L., Shan, J. & Mak, K. F. Ising pairing in superconducting NbSe2 atomic layers. *Nat Phys* **12**, 139-143, (2016).
3. Xing, Y., Zhao, K., Shan, P., Zheng, F., Zhang, Y., Fu, H., Liu, Y., Tian, M., Xi, C., Liu, H., Feng, J., Lin, X., Ji, S., Chen, X., Xue, Q.-K. & Wang, J. Ising Superconductivity and Quantum Phase Transition in Macro-Size Monolayer NbSe2. *Nano Lett* **17**, 6802-6807, (2017).
4. de la Barrera, S. C., Sinko, M. R., Gopalan, D. P., Sivadas, N., Seyler, K. L., Watanabe, K., Taniguchi, T., Tsen, A. W., Xu, X., Xiao, D. & Hunt, B. M. Tuning Ising superconductivity with layer and spin–orbit coupling in two-dimensional transition-metal dichalcogenides. *Nat Commun* **9**, 1427, (2018).
5. Zhou, B. T., Yuan, N. F. Q., Jiang, H.-L. & Law, K. T. Ising superconductivity and Majorana fermions in transition-metal dichalcogenides. *Phys Rev B* **93**, 180501, (2016).
6. Triola, C., Badiane, D. M., Balatsky, A. V. & Rossi, E. General Conditions for Proximity-Induced Odd-Frequency Superconductivity in Two-Dimensional Electronic Systems. *Phys Rev Lett* **116**, 257001, (2016).
7. Hsu, Y.-T., Vaezi, A., Fischer, M. H. & Kim, E.-A. Topological superconductivity in monolayer transition metal dichalcogenides. *Nat Commun* **8**, 14985, (2017).
8. Rahimi, M. A., Moghaddam, A. G., Dykstra, C., Governale, M. & Zülicke, U. Unconventional superconductivity from magnetism in transition-metal dichalcogenides. *Phys Rev B* **95**, 104515, (2017).
9. He, W.-Y., Zhou, B. T., He, J. J., Yuan, N. F. Q., Zhang, T. & Law, K. T. Magnetic field driven nodal topological superconductivity in monolayer transition metal dichalcogenides. *Communications Physics* **1**, 40, (2018).
10. Xi, X., Zhao, L., Wang, Z., Berger, H., Forró, L., Shan, J. & Mak, K. F. Strongly enhanced charge-density-wave order in monolayer NbSe2. *Nat Nanotechnol* **10**, 765, (2015).
11. Bischoff, F., Auwärter, W., Barth, J. V., Schiffrin, A., Fuhrer, M. & Weber, B. Nanoscale Phase Engineering of Niobium Diselenide. *Chem Mater* **29**, 9907-9914, (2017).
12. Nakata, Y., Sugawara, K., Shimizu, R., Okada, Y., Han, P., Hitosugi, T., Ueno, K., Sato, T. & Takahashi, T. Monolayer 1T-NbSe2 as a Mott insulator. *NPG Asia Mater* **8**, e321, (2016).
13. Ugeda, M. M., Bradley, A. J., Zhang, Y., Onishi, S., Chen, Y., Ruan, W., Ojeda-Aristizabal, C., Ryu, H., Edmonds, M. T., Tsai, H.-Z., Riss, A., Mo, S.-K., Lee, D., Zettl, A., Hussain, Z., Shen, Z.-X. & Crommie, M. F. Characterization of collective ground states in single-layer NbSe2. *Nat Phys* **12**, 92, (2015).
14. Mattheiss, L. F. Band Structures of Transition-Metal-Dichalcogenide Layer Compounds. *Phys Rev B* **8**, 3719-3740, (1973).
15. Nakano, M., Wang, Y., Kashiwabara, Y., Matsuoka, H. & Iwasa, Y. Layer-by-Layer Epitaxial Growth of Scalable WSe2 on Sapphire by Molecular Beam Epitaxy. *Nano Lett* **17**, 5595-5599, (2017).

16  Moser, J., Matos-Abiague, A., Schuh, D., Wegscheider, W., Fabian, J. & Weiss, D. Tunneling Anisotropic Magnetoresistance and Spin-Orbit Coupling in Fe/GaAs/Au Tunnel Junctions. *Phys Rev Lett* **99**, 056601, (2007).
17  Žutić, I., Matos-Abiague, A., Scharf, B., Dery, H. & Belashchenko, K. Proximitized materials. *Mater Today* **22**, 85-107, (2019).
18  Samnakay, R., Wickramaratne, D., Pope, T. R., Lake, R. K., Salguero, T. T. & Balandin, A. A. Zone-Folded Phonons and the Commensurate–Incommensurate Charge-Density-Wave Transition in 1T-TaSe2 Thin Films. *Nano Lett* **15**, 2965-2973, (2015).
19  Aytan, E., Debnath, B., Kargar, F., Barlas, Y., Lacerda, M. M., Li, J. X., Lake, R. K., Shi, J. & Balandin, A. A. Spin-phonon coupling in antiferromagnetic nickel oxide. *Appl Phys Lett* **111**, 252402, (2017).
20  Pak, S., Lee, J., Lee, Y.-W., Jang, A. R., Ahn, S., Ma, K. Y., Cho, Y., Hong, J., Lee, S., Jeong, H. Y., Im, H., Shin, H. S., Morris, S. M., Cha, S., Sohn, J. I. & Kim, J. M. Strain-Mediated Interlayer Coupling Effects on the Excitonic Behaviors in an Epitaxially Grown MoS2/WS2 van der Waals Heterobilayer. *Nano Lett* **17**, 5634-5640, (2017).
21  Pereira, C. M. & Liang, W. Y. Raman studies of the normal phase of 2H-NbSe 2. *Journal of Physics C: Solid State Physics* **15**, L991, (1982).
22  Bardeen, J., Cooper, L. N. & Schrieffer, J. R. Theory of Superconductivity. *Phys Rev* **108**, 1175-1204, (1957).
23  Fu, Y., Liu, E., Yuan, H., Tang, P., Lian, B., Xu, G., Zeng, J., Chen, Z., Wang, Y., Zhou, W., Xu, K., Gao, A., Pan, C., Wang, M., Wang, B., Zhang, S.-C., Cui, Y., Hwang, H. Y. & Miao, F. Gated tuned superconductivity and phonon softening in monolayer and bilayer MoS2. *npj Quantum Materials* **2**, 52, (2017).
24  Hui, Y. Y., Liu, X., Jie, W., Chan, N. Y., Hao, J., Hsu, Y.-T., Li, L.-J., Guo, W. & Lau, S. P. Exceptional Tunability of Band Energy in a Compressively Strained Trilayer MoS2 Sheet. *Acs Nano* **7**, 7126-7131, (2013).
25  Staley, N. E., Wu, J., Eklund, P., Liu, Y., Li, L. & Xu, Z. Electric field effect on superconductivity in atomically thin flakes of NbSe2. *Phys Rev B* **80**, 184505, (2009).
26  El-Bana, M. S., Wolverson, D., Russo, S., Balakrishnan, G., Paul, D. M. & Bending, S. J. Superconductivity in two-dimensional NbSe2 field effect transistors. *Superconductor Science and Technology* **26**, 125020, (2013).
27  Xi, X., Berger, H., Forró, L., Shan, J. & Mak, K. F. Gate Tuning of Electronic Phase Transitions in Two-Dimensional NbSe2. *Phys Rev Lett* **117**, 106801, (2016).
28  Ye, J. T., Zhang, Y. J., Akashi, R., Bahramy, M. S., Arita, R. & Iwasa, Y. Superconducting Dome in a Gate-Tuned Band Insulator. *Science* **338**, 1193, (2012).
29  Zhou, T., Dartiailh, M. C., Mayer, W., Han, J. E., Matos-Abiague, A., Shabani, J. & Žutić, I. Phase Control of Majorana Bound States in a Topological X Junction. *Phys Rev Lett* **124**, 137001, (2020).
30  Wang, Z., Tang, C., Sachs, R., Barlas, Y. & Shi, J. Proximity-Induced Ferromagnetism in Graphene Revealed by the Anomalous Hall Effect. *Phys Rev Lett* **114**, 016603, (2015).
31  Wei, P., Lee, S., Lemaitre, F., Pinel, L., Cutaia, D., Cha, W., Katmis, F., Zhu, Y., Heiman, D., Hone, J., Moodera, J. S. & Chen, C.-T. Strong interfacial exchange field in the graphene/EuS heterostructure. *Nat Mater* **15**, 711, (2016).
32  Perdew, J. P., Burke, K. & Ernzerhof, M. Generalized Gradient Approximation Made Simple. *Phys Rev Lett* **77**, 3865-3868, (1996).

33  Kresse, G. & Joubert, D. From ultrasoft pseudopotentials to the projector augmented-wave method. *Phys Rev B* **59**, 1758-1775, (1999).
34  Grimme, S. Semiempirical GGA-type density functional constructed with a long-range dispersion correction. *J Comput Chem* **27**, 1787-1799, (2006).
35  Towns, J., Cockerill, T., Dahan, M., Foster, I., Gaither, K., Grimshaw, A., Hazlewood, V., Lathrop, S., Lifka, D., Peterson, G. D., Roskies, R., Scott, J. R. & Wilkins-Diehr, N. XSEDE: Accelerating Scientific Discovery. *Comput Sci Eng* **16**, 62-74, (2014).

# Supporting Information:

# Strain controlled superconductivity in few-layer NbSe$_2$

Cliff Chen[1], Protik Das[2], Ece Aytan[2], Weimin Zhou[1], Justin Horowitz[1], Biswarup Satpati[3], Alexander A. Balandin[2], Roger Lake[2] and Peng Wei[1]*

1. Department of Physics and Astronomy, University of California, Riverside, CA 92521, United States

2. Department of Electrical and Computer Engineering, University of California, Riverside, CA 92521, United States

3. Surface Physics & Material Science Division, Saha Institute of Nuclear Physics, 1/AF Bidhannagar, Kolkata 700 064, India

*Correspondence to: peng.wei@ucr.edu

## S1. Sample growth:

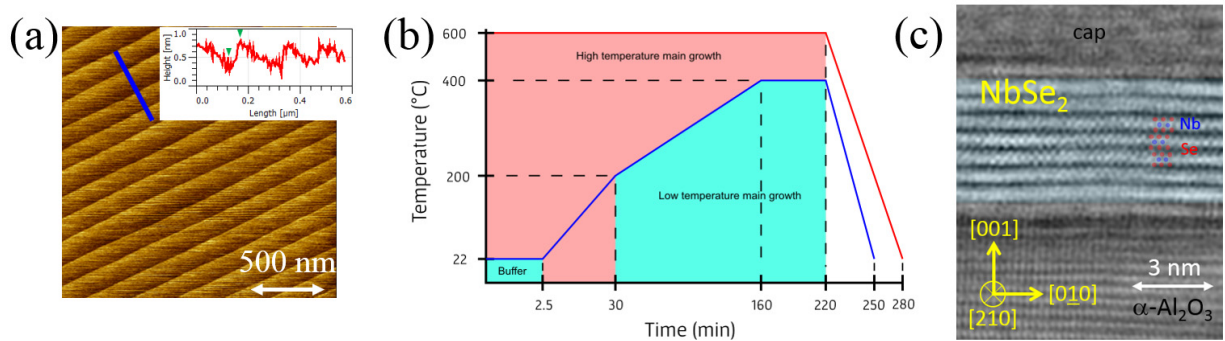

**Figure S1**: (a) The atomic force microscopy (AFM) image of the sapphire (0001) substrate that we used for growing wafer-scale NbSe$_2$. The inset show the height of the atomic terraces. (b) Outline of NbSe$_2$ thin film growth. Red and blue curves are for high temperature and low temperature growths respectively. Color shaded areas indicate intervals when the Nb shutter is open; Se shutter is kept open throughout entire growth. Beam interruption of the Nb flux is used in the low temperature main growth. (c) The atomically resolved scanning transmission electron microscopy (STEM) on a 7ML NbSe$_2$ sample. The relative arrangement of Nb and Se atoms is clearly observed

The atomic terrace of the sapphire substrate (Fig. S1a) is obtained by annealing the substrate at 1200 °C in ambient environment for 5 hours. Prior to deposition, the sapphire substrates were degassed in-situ at 600℃ for 30 minutes to remove surface adsorbents before cooling to the appropriate temperature to start the growth. The Nb and Se sources were heated via an electron beam evaporator and Knudsen cell, respectively. Fig. S1b outlines two types of growth procedures employed, one at low temperatures and the other one at high temperatures. For the low temperature growth, an amorphous buffer layer of Nb and Se is first deposited onto the substrate at room temperature in a similar manner as described in Ref [1] with a thickness of about half a monolayer. The buffer is subsequently annealed to 200℃ at which point the main growth begins. We employ a beam interrupted method during the main growth in which the Nb shutter is cyclically closed and opened (10s open, 30s closed) to prevent excess Nb in the film as well as to promote selenization of the Nb atoms. Meanwhile, the substrate heating is steadily raised to its final temperature of 400℃. After growth, the substrate is cooled to room temperature when it is capped with 30nm of Se. It should be noted that the Se flux is supplied continuously throughout the entire growth. As for the high temperature growth, we did not deposit a buffer layer.

**S2. Sample characterization:**

The spatial locations of the Nb and Se atoms is visualized by atomically resolved scanning transmission electron microscopy (STEM) as shown in Fig S1c. The relative arrangement suggests 2H phase. The spatially revolved dispersive X-ray spectroscopy (EDX) is taken at the interface of the layer (Fig S2b). No contamination is found, and there is also no phase segregation is found in the layer. The Nb:Se ~ 1:2 ratio is determined by integrating the spectra peaks in Fig S2b.

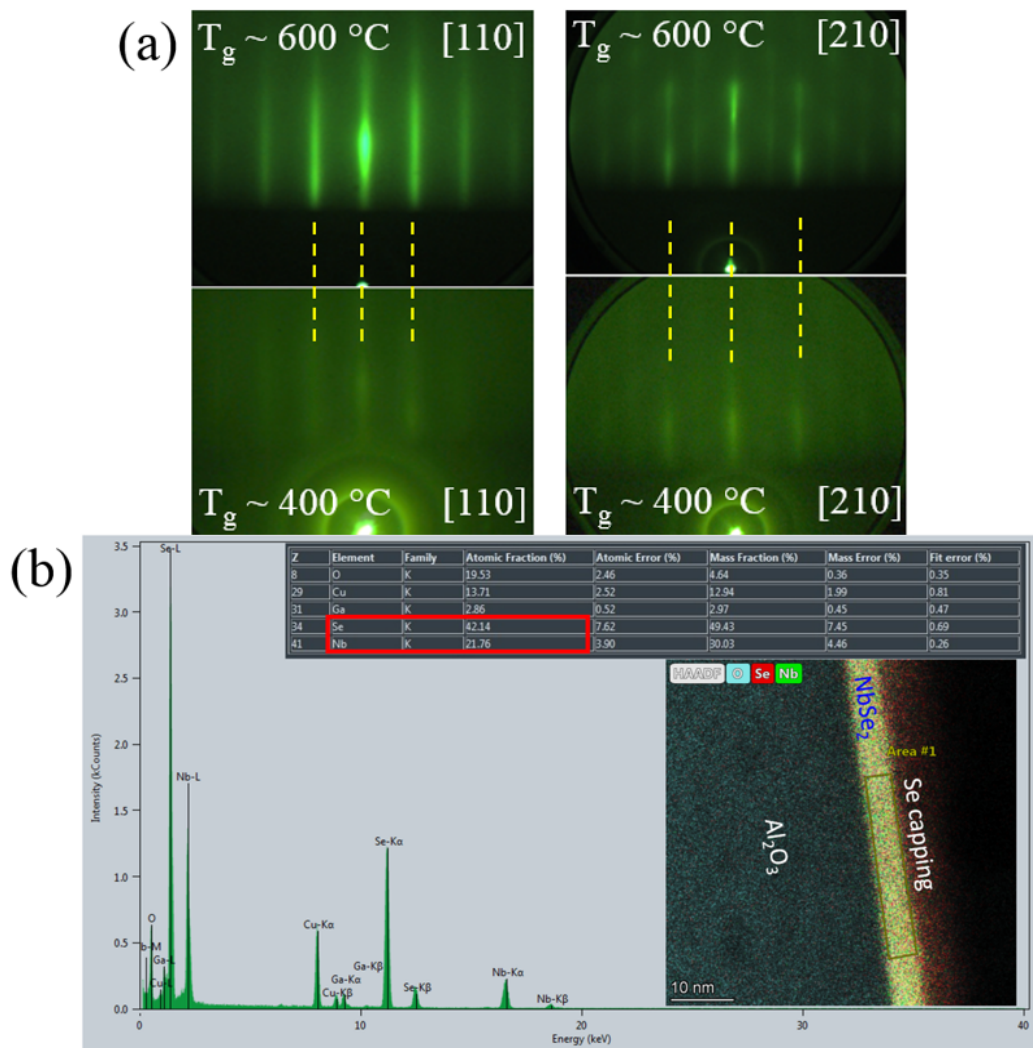

**Figure S2**: (a) The comparison of RHEED diffraction patterns of the insulating and superconducting NbSe$_2$ samples. They both have similar six-fold planar rotation symmetry and comparable lattice constants. The legend on the figure labels the sample growth temperature $T_g$ and the incidence direction of the RHEED electron beam. It is clear that for the sample grown at $T_g \sim 600$ °C the RHEED streaks are much sharper, which confirms the higher quality for the sample grown at higher temperature. (b) EDX data taken at a cross section of the NbSe$_2$ layer (right bottom). Nb and Se atoms are uniformly distributed in the layer indicating no phase segregations,

and also no interdiffusion of atoms in between the layer and the substrate. The spectrum analysis shows the ratio of Nb:Se ~ 1:2.

Our experiment shows that for NbSe2 grown at 600 °C the samples have a better quality. This is consistent to the common knowledge of NbSe$_2$ growth.[2] The surface quality of our sample is monitored in-situ during the growth using reflection high energy electron diffraction (RHEED). Fig. S2a shows the results of RHEED in two samples grown at two different temperatures. The bright diffraction lines (often called streaks) is a constructive interference of electrons after being scattered by the crystal lattice. Narrower and sharper streaks indicate better crystal quality. It is clear that the sample grown at $T_g$ ~ 600 °C has better quality compared to the sample grown at $T_g$ ~ 400 °C. Also, both samples have the same six-fold rotation symmetry.

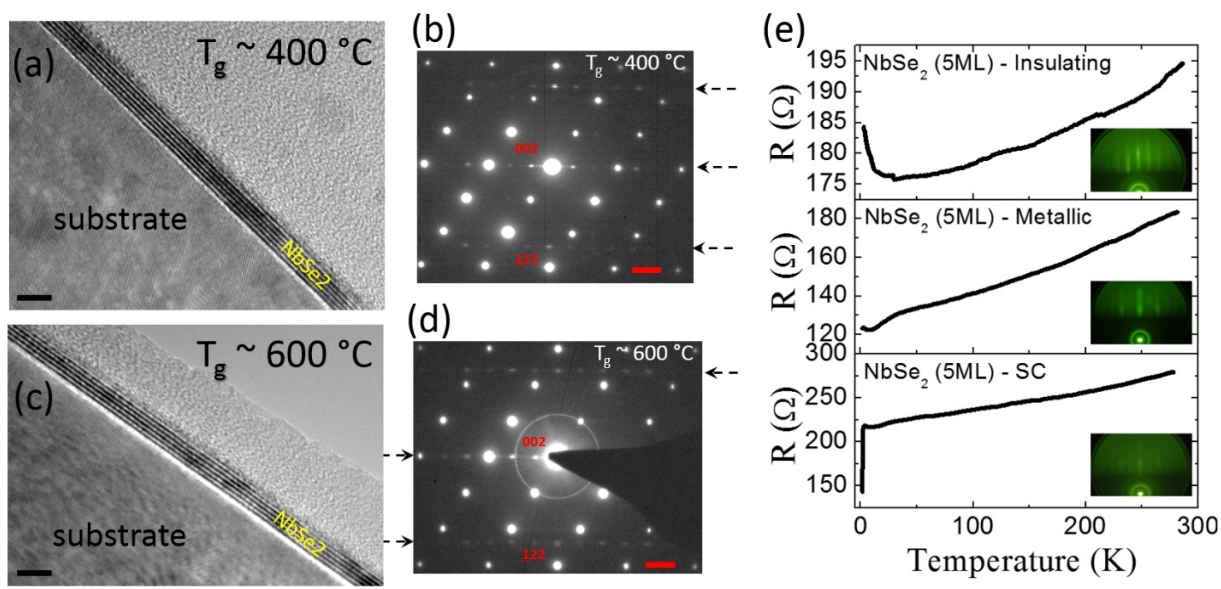

Figure S3: The comparison of the TEM image (a) (c) and SAD data (b) (d) among two samples grown at different growth temperatures $T_g$ ~ 400 °C and $T_g$ ~ 600 °C. It can be clearly seen that for both samples the monolayer of five-layer NbSe$_2$ are well defined (a, c). Both samples have a sharp interface with the substrate. The black line marks in (a, c) indicate 5 nm. For the SAD diffraction data (b, d), which is taken by performing electron diffraction across the interface, the sample grown at $T_g \sim 600$ °C has better defined diffraction spots. The location of the lines for the diffraction spots of NbSe$_2$ are indicated by the dotted arrows in (b) and (d). Sharper diffractions spots can be seen in (d) and more spots are visible in (d). It thus indicates that the sample grown at $T_g \sim 600$ °C has higher quality. (e) The comparison of the resistance among three control samples of NbSe$_2$ grown side-by-side. The samples are kept similar to each other with exactly the same thickness (5 ML) and are grown on similar substrates except that the growth conditions such as the growth temperatures are varied. Their resistance values are comparable. The inset shows the corresponding RHEED diffraction pattern of each sample.

The interface quality of the sample has been examined using TEM and cross-sectional SAD. Fig. S3 shows the results of TEM and SAD among two samples grown at two different temperatures. TEM can visualize the deterioration of the sample. However, comparing Fig S3a and Fig S3c, no deterioration of the sample can be seen. Both samples have clearly defined NbSe$_2$ atomic planes and well-defined interface. Further, by comparing Fig S3b and Fig S3d, we can see that the SAD pattern is sharper for the sample grown at $T_g \sim 600$ °C. It also proves that the sample grown at 600 °C has a better quality.

To compare the actual resistance values of the samples, we have grown a batch of NbSe$_2$ with the thickness fixed exactly at 5 ML and measured their resistances in the same measurement geometry. The samples are grown on similar substrates prepared side-by-side following similar growth procedures except that the growth conditions such as the growth temperatures are varied. As can be seen (Fig S3e), three samples demonstrate a transition from superconducting to insulating behavior. However, their resistance values are comparable. Especially, for the insulating sample, its resistance is even slightly smaller than the superconducting sample. This is in agreement with the improved crystalline quality in the insulating sample as can be seen from the sharper RHEED diffraction patterns in the inset. The comparable resistance values also suggest that there is no dramatic reduction on the carrier density and on the crystalline quality in the insulating NbSe2 sample.

**S3. Variable range hopping fitting to the insulating sample:**

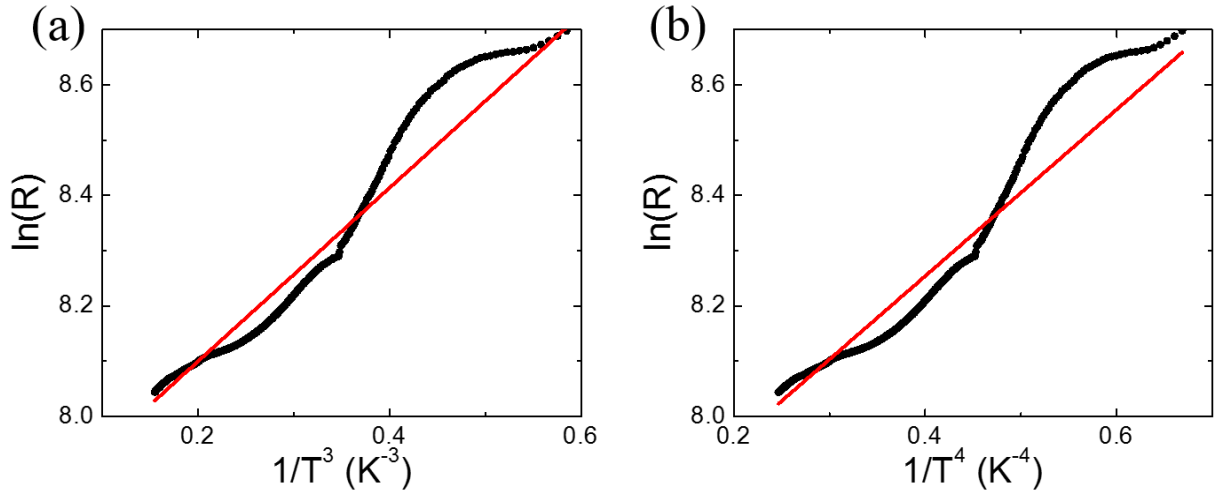

**Figure S4**: (a) The two-dimensional VRH fitting to the insulating NbSe2 sample. (b) The three-dimensional VRH fitting. Both of them do not describe the temperature dependence of the resistance.

To clarify that the insulating behavior of the NbSe2 sample grown at high temperature (main text Fig. 2a blue) is not due to localization effect as a result of degradation of the sample quality, we compared the fitting using variable range hopping (VRH) model in both two-dimension and three-dimension cases.[3] As can be seen in Fig. S1, both of the plots $\ln(R)$ vs. $T^{-3}$ (2D case) and $\ln(R)$ vs. $T^{-4}$ (3D case) could not be fitted by a line. In contrast, Fig. 3c in the main text best describe $\sigma(T)$ as $\sigma \sim e^{-\frac{E_0}{k_B T}}$, which indicates that the insulating behavior is due to an energy gap.

## S4. DFT calculations of the phonon spectrum of 1T-NbSe$_2$:

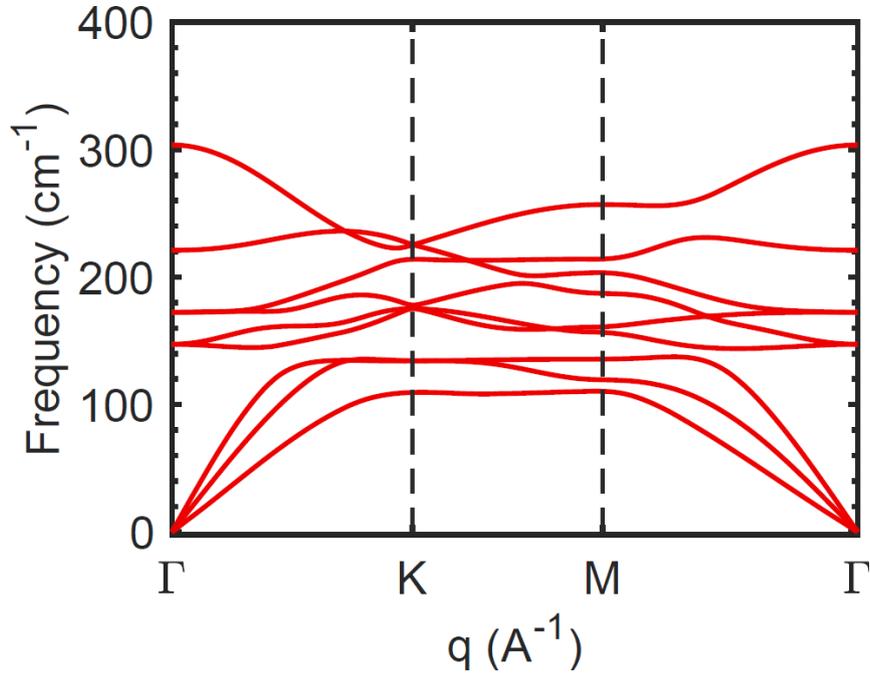

**Figure S5**: The phonon spectrum of 1T-NbSe$_2$ calculated by DFT. There are four Raman active modes below 300 cm$^{-1}$.

While optimizing the electronic structures, the vdW interactions were accounted for using a semi-empirical correction to the Kohn Sham energies also known as Grimme's DFT-D2 approach.[4] For all the electronic structure calculations, a Monkhorst−Pack scheme was used to integrate over the Brillouin zone with a k-mesh of $8 \times 8 \times 8$. A plane-wave basis cutoff of $400 \ eV$ was used for all the calculations. All structures were optimized using the BFGS optimizer implemented in the Python package atomic simulation environment (ASE).[5] VASP was used as a force calculator for the optimization. The structures were optimized until the largest force on the atoms was less than $10^{-6}$ eV/Å. The optimized lattice constants for bulk 1T-NbSe$_2$ were $a = 3.60$ Å and $c = 6.36$ Å. ASE was used to programmatically optimize all of the structures as a function of strain. During the optimization, strain was applied equally to lattice constants $a$ and $b$ and kept fixed throughout the optimization. The lattice constant $c$ was allowed to relax. During the cell optimization under strain, the cell shape was also allowed to change, but it remained unchanged. The small displacement method implemented in software package Phonopy was used to calculate the phonon dispersion of the optimized structures.[6] A converged supercell size of $2 \times 2 \times 2$ was used for all the phonon dispersion calculations.

**References:**

1   Nakano, M., Wang, Y., Kashiwabara, Y., Matsuoka, H. & Iwasa, Y. Layer-by-Layer Epitaxial Growth of Scalable WSe2 on Sapphire by Molecular Beam Epitaxy. *Nano Lett* **17**, 5595-5599, (2017).
2   Cheng, F., Ding, Z., Xu, H., Tan, S. J. R., Abdelwahab, I., Su, J., Zhou, P., Martin, J. & Loh, K. P. Epitaxial Growth of Single-Layer Niobium Selenides with Controlled Stoichiometric Phases. *Advanced Materials Interfaces* **5**, 1800429, (2018).
3   Mott, N. F. Conduction in non-crystalline materials. *The Philosophical Magazine: A Journal of Theoretical Experimental and Applied Physics* **19**, 835-852, (1969).
4   Grimme, S. Semiempirical GGA-type density functional constructed with a long-range dispersion correction. *J Comput Chem* **27**, 1787-1799, (2006).
5   Hjorth Larsen, A., Jørgen Mortensen, J., Blomqvist, J., Castelli, I. E., Christensen, R., Dułak, M., Friis, J., Groves, M. N., Hammer, B., Hargus, C., Hermes, E. D., Jennings, P. C., Bjerre Jensen, P., Kermode, J., Kitchin, J. R., Leonhard Kolsbjerg, E., Kubal, J., Kaasbjerg, K., Lysgaard, S., Bergmann Maronsson, J., Maxson, T., Olsen, T., Pastewka, L., Peterson, A., Rostgaard, C., Schiøtz, J., Schütt, O., Strange, M., Thygesen, K. S., Vegge, T. *et al.* The atomic simulation environment—a Python library for working with atoms. *Journal of Physics: Condensed Matter* **29**, 273002, (2017).
6   Togo, A. & Tanaka, I. First principles phonon calculations in materials science. *Scripta Mater* **108**, 1-5, (2015).